\documentclass[summary]{URSIGASS2021}

\usepackage{siunitx}
\usepackage{subfigure}
\usepackage{caption}
\usepackage{amsmath, amssymb}
\usepackage{bigints}
\usepackage{tikz,pgfplots}
\usepackage{tikz}
\usepackage{booktabs}
\usetikzlibrary{shapes,arrows,shapes.multipart,positioning,decorations.text,fit}
\usepackage{tikz,pgfplots}
\usepgfplotslibrary{colormaps}

\title{Target Characteristics on Rotation in Euclidean Space Using Full Polarimetric SAR Data}

\author{{Subhadip~Dey}\affref{ref1}\affref{ref2}, and {Irena~Hajnsek}\affref{ref1}}

\affiliation{%
  \aff{ref1}{Microwaves and Radar Institute, German Aerospace Center (DLR), We\ss ling, Germany}
\aff{ref2}{Agricultural and Food Engineering Department,
Indian Institute of Technology Kharagpur, India}
  }

\begin{document}

\maketitle

\pgfplotsset{
    colormap/jet,
}
\begin{abstract}
In radar polarimetry, the target characteristics significantly depend on the rotation of the target concerning the radar line of sight. Hence, several attempts have been made in the literature to compensate for the rotational dependency or to derive a complete roll-invariant parameter using full polarimetric SAR data. However, the degree of dependency of the targets on the rotation domain has yet to be well explored. Hence, in this study, we have proposed a new parameter that characterizes the targets concerning rotation in Euclidean space. The parameter shows similar values for the targets, which can be related to each other through a unitary transformation. The advantage of this parameter has been demonstrated over different canonical targets. Further, the characteristics of the natural targets have been shown using the Single Look Complex data of C band Radarsat-2 and ALOS PALSAR over the San Francisco Bay area and Cuba, respectively.
\end{abstract}

\section{Introduction}
In radar target phenomenology, the rotation of the target, about the radar line of sight, plays a significant role in determining the target type. In this regard, full polarimetric Synthetic Aperture Radar (SAR) data can be used to determine a target's rotation angle. As a result, different applications of rotation in Euclidean space, such as model-based decomposition techniques and the computation of roll-invariant parameters, are performed using full-polarimetric SAR data.  

As mentioned earlier, one of the general interests in polarimetric SAR is determining the rotation angle to remove the orientation-induced polarimetric distortion. In particular, the compensation of the target rotation has been performed by maximizing the co-polar response~\cite{schuler1996measurement}. Lee et al.~\cite{lee2000polarimetric} utilized the left-hand and right-hand circular polarization vectors to estimate the rotation angle from single-look and multilook images. However, it was also mentioned that the polarization signature provides a better approximation than the circular polarization vector-based method. At the same time, the same authors proposed a unified analysis of orientation shifts induced by terrain slopes using the circular polarization covariance matrix~\cite{lee2002estimation}. Therefore, it can be understood from these algorithms that in the absence of the helical component, the rotation can be approximated by the phase of a pair of circular polarizations or by the elements of the covariance matrix. Similarly, in the absence of helicity, Cloude-Pottier parameterization of the eigen component can be used to estimate the target rotation~\cite{cloude2009polarisation}. However, when the helicity exists within the resolution cell, the target scattering decomposition proposed by Touzi~\cite{touzi2006target} can be used, where a new scattering vector model has been proposed by projecting the Kennaugh-Huynen con-diagonalization into Pauli basis.

The concept of the target rotation has also been applied in the model-based decomposition techniques. A rotation of the coherency matrix was applied to minimize the effect of the cross-polarized component before performing the four-component decomposition~\cite{yamaguchi2011four}. This application of the minimization of the cross-pol component using the rotation matrix enhanced the separability between the oriented urban area and vegetation structures. Similarly, Xu and Jin~\cite{xu2005deorientation} deoriented the polarimetric scattering information to obtain a new set of scattering parameters, including a deorientation angle, $\psi$. An et al.~\cite{an2010three} utilized the rotation of the Huynen parameters to maximize the co-polarized power component.

Apart from the applications of rotation on the scattering information, an effort has also been made to derive parameters independent of the target rotation concerning the radar line of sight. For example, a scattering-type parameter, $\alpha$, was proposed to distinguish diverse scattering targets in both coherent and incoherent domains~\cite {cloude2009polarisation}. However, $\alpha$ cannot separate a dihedral target from a helix target as $\alpha = \ang{90}$ for both targets. Hence, Touzi~\cite{touzi2006target} proposed two new scattering parameters, $\alpha_s$ and $\Phi_{\alpha_s}$ to overcome the ambiguity between those targets. On the other hand, Dey et al.~\cite{dey2020target} proposed another new target characterization parameter, $\theta_{\text{FP}}$, which enhanced the overall separation capability of different land cover targets. $\theta_{\text{FP}}$ has also been widely used in other SAR applications~\cite{dey2021model, dey2020novel, dey2022polarimetric}. Recently, Chen et al.~\cite{chen2013uniform} showed different roll-invariant parameters using the elements of the Sinclair, coherency and covariance matrices.

Notably, the existing studies either try to find a rotation angle to minimize/ maximize the cross-/ co-pol scattering power components or to develop roll-invariant parameters. However, in particular, the overall oscillation of a scattering target due to different rotation angles is overlooked. Hence, in this study, we propose a new parameter which will provide insight into the degree of oscillation of a scattering target depending on the complete rotation spectrum in Euclidean space. The proposed parameter is interpreted using the full polarimetric Single Look Complex (SLC) data of C band Radarsat-2 over the San Francisco Bay area and L band ALOS PALSAR over Cuba.

\section{Methodology}
In full polarimetric radar data the scattering information can be represented in terms of the Sinclair matrix, $\mathbf{S}$,

\begin{equation}
    \mathbf{S} = \begin{bmatrix}
        S_{HH} & S_{HV}\\
        S_{VH} & S_{VV}
    \end{bmatrix}
\end{equation}

where, $S_{HV}$ is the scattering information of the horizontal transmit and vertical received signal. Other terms can be defined in a similar way. When the matrix is rotated at an angle $\theta$ with respect to the radar line of sight, the rotated matrix becomes,

\begin{equation}
    \mathbf{S}(\theta) = \mathbf{R}(\theta)~\mathbf{S}~\mathbf{R}^{*T}(\theta)
\end{equation}

where, ${*T}$ denotes the conjugate transpose, and $\mathbf{R}(\theta)$ is a $2 \times 2$ rotation matrix, which can be defined as, $\mathbf{R}(\theta) = \begin{bmatrix}
    \cos{\theta} & -\sin{\theta}\\
    \sin{\theta} & \cos{\theta}
\end{bmatrix}$.

where, $\theta \in \left[\ang{0}, \ang{180}\right]$. To obtain the realizations of $S_{HH},~S_{HV}$ and $S_{VV}$ within the range of $\theta$, we discretize the range at an interval of $\ang{1}$. Thereafter we calculate the mean values of $|S_{HH}|,~|S_{HV}|$ and $|S_{VV}|$ as,

\begin{equation}
    \overline{S}_{XY} = \dfrac{\sum_{\theta = \ang{0}}^{\ang{180}} |S_{XY} (\theta)|}{N} 
\end{equation}

where, $X$ and $Y$ denote $H$ or $V$ polarizations, $N$ denotes the total number of realizations which in this case $= \SI{181}{}$ and $|\cdot|$ represents the amplitude. Further, we normalize the mean values of $|S_{HH}|,~|S_{HV}|$ and $|S_{VV}|$ as,

\begin{equation}
    \widehat{S}_{XY} = \dfrac{\overline{S}_{XY}}{\overline{S}_{HH} + \overline{S}_{HV} + \overline{S}_{VV}}
\end{equation}

where, $\widehat{S}_{XY}$ is the normalized value of $\overline{S}_{XY}$ which varies between 0 and 1. This normalized element is then represented in terms of the cosine as, $\phi_{XY} = \cos^{-1}(\widehat{S}_{XY})$. In this way we obtain three angular elements, such as, $\phi_{HH}$, $\phi_{HV}$, and $\phi_{VV}$ which vary from \ang{0} to \ang{90}.

In addition to the mean values of $|S_{HH}|,~|S_{HV}|$ and $|S_{VV}|$, we also obtain the standard deviations, $\sigma_{HH}$, $\sigma_{HV}$ and $\sigma_{VV}$ as,

\begin{equation}
    \sigma_{XY} = \sqrt{\dfrac{\sum_{\theta = \ang{0}}^{\ang{180}} (|S_{XY} \left(\theta)| - \overline{S}_{XY}\right)^2}{N}} 
\end{equation}

The obtained standard deviation values are then normalized and a new parameter, $\zeta$ is derived which infers the overall oscillation of a scattering target within the rotation domain. 

\begin{equation}
    \widehat{\sigma}_{XY} = \dfrac{\sigma_{XY}}{\sigma_{HH} + \sigma_{HV} + \sigma_{VV}}
\end{equation}

\begin{equation}
    \zeta = \widehat{\sigma}_{HH} \times \phi_{HH} + \widehat{\sigma}_{HV} \times \phi_{HV} + \widehat{\sigma}_{VV} \times \phi_{VV}
\end{equation}

This $\zeta$ parameter is used to describe the oscillation due to rotation in Euclidean space for different canonical, natural and human-made targets.

\section{Results and Discussion}
In this section we have first analyze the characteristics of $\zeta$ for different canonical targets and then we analyze over the natural and human-made targets using the Radarsat-2 data.

\subsection{Analysis over canonical targets}
Table~\ref{tab:zeta_canonical} shows the $\zeta$ value for Trihedral, dihedral, cross-pol, helix, horizontal dipole and \ang{60} oriented dipole targets. It can be seen that $\zeta$ over Trihedral lies at \ang{0}, indicating no dependency on the rotation in the Euclidean space, while the dihedral target shows a $\zeta$ value of \ang{64.9}. However, it is exciting to note that the $\zeta$ value of cross-pol is also at \ang{64.9}. The similarity of $\zeta$ values between dihedral and cross-pol is because of the direct relation through unitary transformation. If the $\mathbf{S}$ matrix of the dihedral target is rotated at \ang{45}, the rotated $\mathbf{S}$ matrix becomes the cross-pol. Therefore, these two targets are not unique in $\zeta$ space. 

\begin{table}[hbt]
    \centering
    \caption{Values of $\zeta$ for different canonical targets}
    \begin{tabular}{lcc}\toprule
        \textbf{Canonical Targets} & $\mathbf{S}$ & $\zeta$\\ \midrule
        Trihedral & $\begin{bmatrix}
            1 & 0\\
            0 & 1
        \end{bmatrix}$ & \ang{0}\\
        Dihedral & $\begin{bmatrix}
            1 & 0\\
            0 & -1
        \end{bmatrix}$ & \ang{64.9}\\
        Cross-pol & $\begin{bmatrix}
            0 & 1\\
            1 & 0
        \end{bmatrix}$ & \ang{64.9}\\
        Helix & $\dfrac{1}{2}\begin{bmatrix}
            1 & i\\
            i & -1
        \end{bmatrix}$ & \ang{23.5}\\
        Horizontal dipole & $\begin{bmatrix}
            1 & 0\\
            0 & 0
        \end{bmatrix}$ & \ang{64.8}\\
        \ang{60} oriented dipole & $\begin{bmatrix}
            0.25 & 0.43\\
            0.43 & 0.75
        \end{bmatrix}$ & \ang{64.8}\\\bottomrule
    \end{tabular}
    \label{tab:zeta_canonical}
\end{table}

A similar analogy can also be drawn between the horizontal dipole and \ang{60} oriented dipole. As both targets are related by a unitary rotation matrix, the $\zeta$ value for both targets is \ang{64.8}. Also, a minor difference of $\zeta$ is observed between dihedral and dipoles. On the contrary, the $\zeta$ value for helix is \ang{23.5}. A possible interpretation of the helix's lower $\zeta$ value compared to the dihedral could be the inherent circularity during scattering, which significantly dampens the dependency on the Euclidean rotation space.

\subsection{Analysis using the Radarsat-2 data}
\begin{figure}[!h]
    \centering
    \subfigure[Pauli RGB]
    {
        \includegraphics[width=0.38\columnwidth]{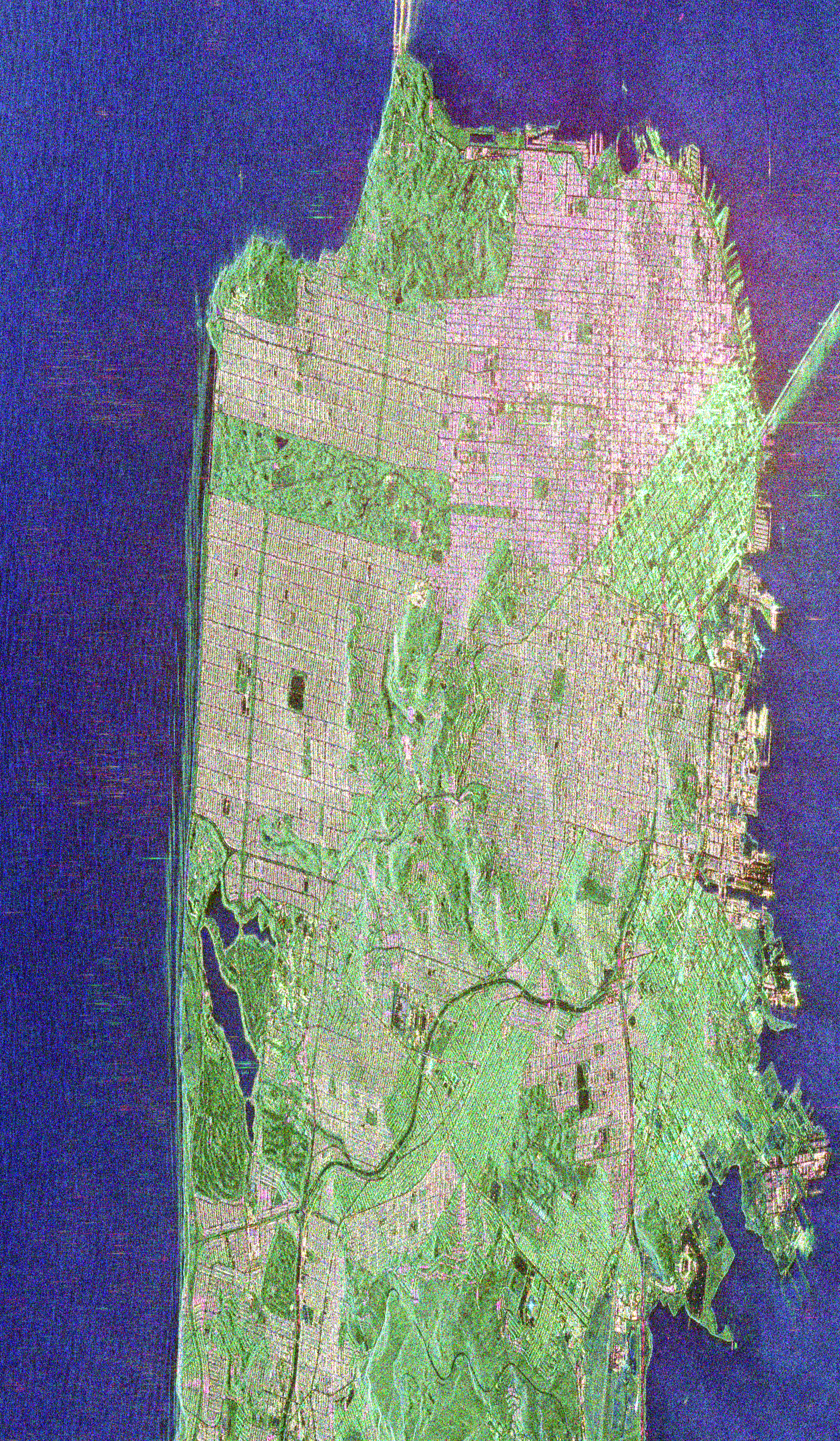}
        \label{fig:pauli_rgb}
    }
    \subfigure[$\zeta$]
    {
        \includegraphics[width=0.49\columnwidth]{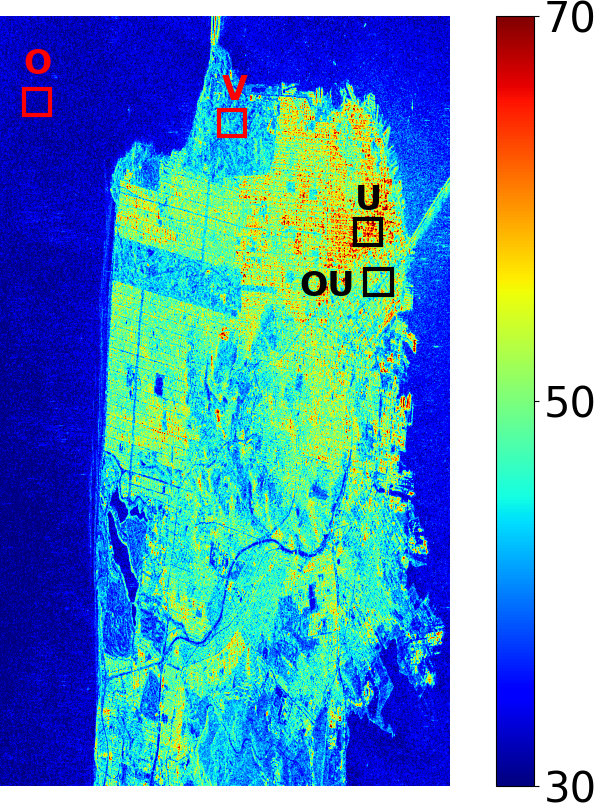}
        \label{fig:zeta_SF}
    }    
    \caption{The Pauli RGB image and the $\zeta$ image over the San Francisco Bay area using the Radarsat-2 data.}
    \label{fig:pauli_zeta_SF}
\end{figure}

Figure~\ref{fig:pauli_zeta_SF} shows the Pauli RGB image and the $\zeta$ image over the San Francisco Bay area using the C-band SLC Radarsat-2 image. Interesting variations of $\zeta$ over different land cover types can be observed. We have kept the range of the colourspace within \ang{30} to \ang{70} as most of the scatterers fall within this range. A nearly uniform value of $\zeta$ is observed over the ocean area, while a distinct range of values is evident over the urban and vegetation zones.  

\begin{figure}[hbt]
    \centering
    \includegraphics[width=\columnwidth]{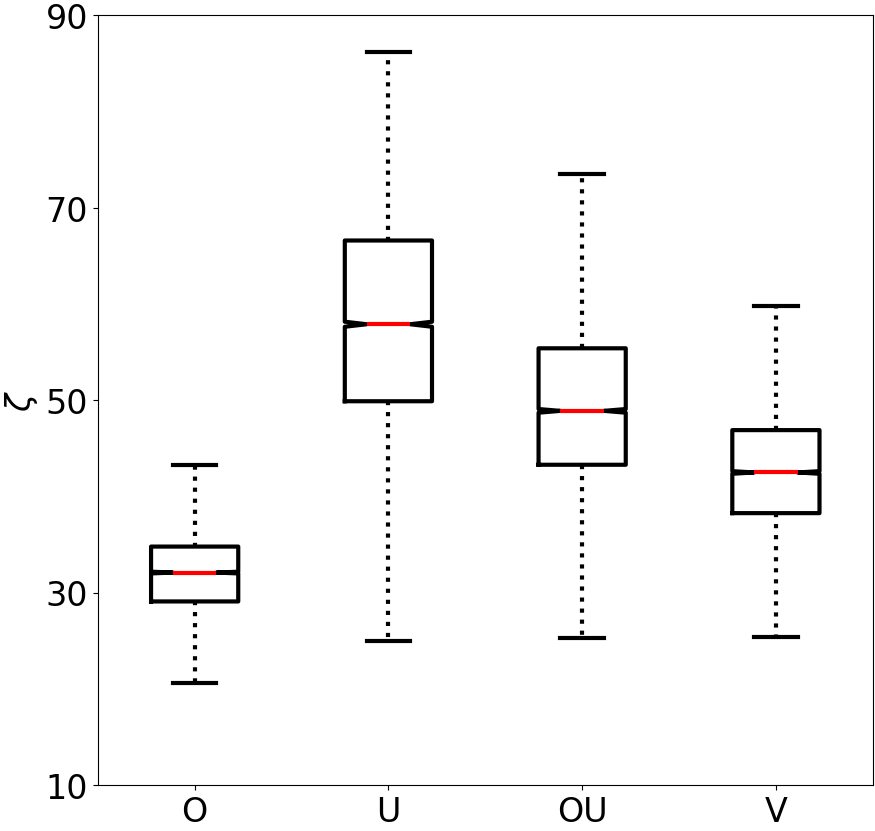}
    \caption{Box plots over the Ocean (`O'), Urban (`U'), Oriented Urban (`OU') and Vegetation (`V') areas using Radarsat-2 data over San Francisco Bay.}
    \label{fig:box_plot_zeta}
\end{figure}

To better understand the characteristics of $\zeta$ over the land cover types we have taken samples over Ocean (`O'), Urban (`U'), Oriented Urban (`OU') and Vegetation (`V') areas as shown in Figure~\ref{fig:zeta_SF}. We then computed the complete variation using the box plot in Figure~\ref{fig:box_plot_zeta}. 

It can be seen from Figure~\ref{fig:box_plot_zeta} that the median value of $\zeta$ over `O' is around \ang{31.2}, while the total range varies around \ang{22} to \ang{47}. As already stated, the variation of $\zeta$ over `O' is nearly steady; hence, the standard deviation of $\zeta$ is also low. Over `U', the median value of $\zeta$ lies around \ang{60}, while over `OU' the median value of $\zeta$ is $\approx$ \ang{50}. In this case, the lower value of $\zeta$ over `OU' as compared to `U' might be due to the existence of the helix scattering~\cite{yamaguchi2011four}, which has dampened the dependency on rotation to a certain extent. In addition to this, it can also be noted that the standard deviation of $\zeta$ is higher for `O' as compared to `OU'.     

\subsection{Analysis using the ALOS PALSAR data}
\begin{figure}[!h]
    \centering
    \subfigure[Pauli RGB]
    {
        \includegraphics[width=0.34\columnwidth]{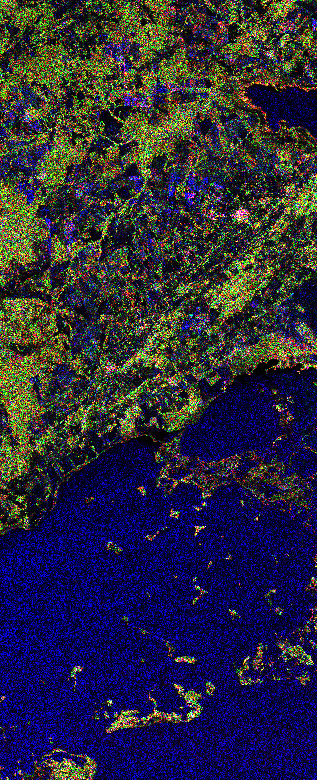}
        \label{fig:pauli_rgb_ALOS}
    }
    \subfigure[$\zeta$]
    {
        \includegraphics[width=0.49\columnwidth]{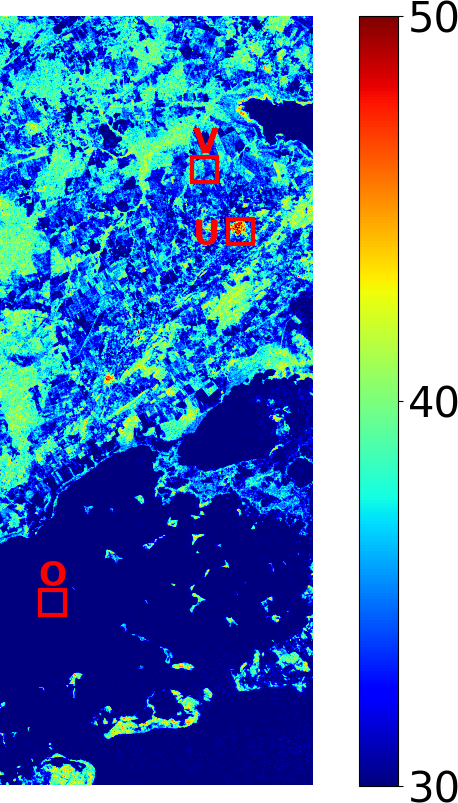}
        \label{fig:zeta_ALOS}
    }    
    \caption{The Pauli RGB image and the $\zeta$ image over Cuba using the ALOS PALSAR data.}
    \label{fig:pauli_zeta_ALOS}
\end{figure}

A similar characteristics of $\zeta$ over land cover types can be seen using L-band ALOS PALSAR data over Cuba in Figure~\ref{fig:pauli_zeta_ALOS}. However, the dynamic range of $\zeta$ gets reduced and in this case we have kept the range of colorspace within \ang{30} to \ang{50}. 

\begin{figure}[hbt]
    \centering
    \includegraphics[width=\columnwidth]{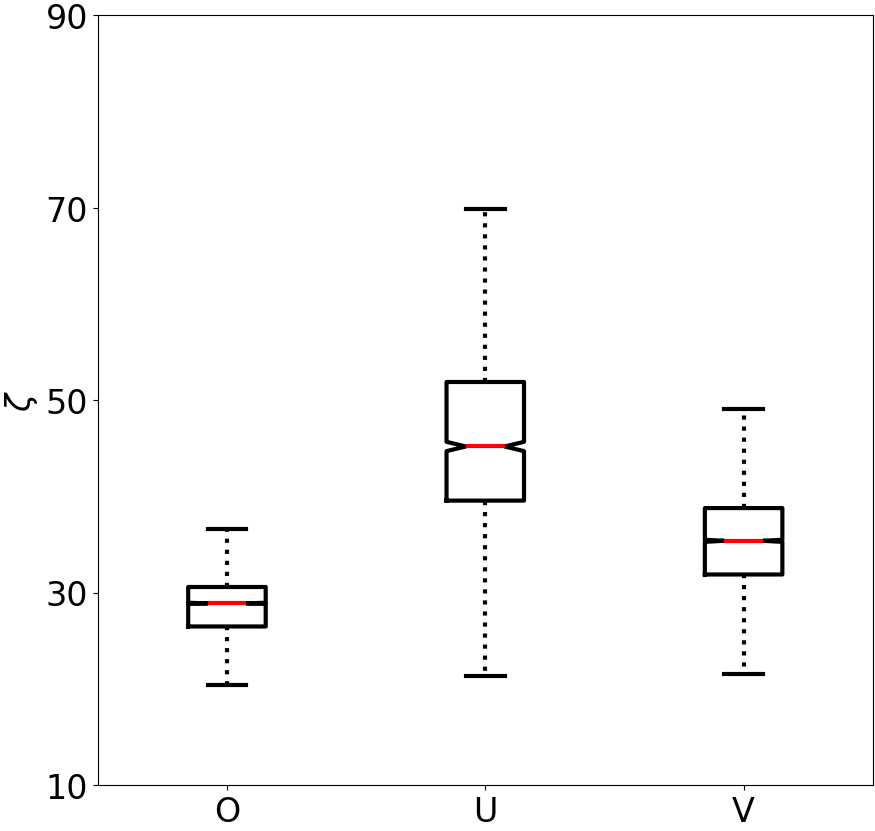}
    \caption{Box plots over the Ocean (`O'), Urban (`U') and Vegetation (`V') areas using ALOS PALSAR data over Cuba.}
    \label{fig:box_plot_zeta_L_band}
\end{figure}

Similar to `C' band data, we observe uniform pattern of $\zeta$ over the ocean surface. However, over the urban area we observe slightly lower value of $\zeta$ as compared to the C-band data. The box plots over `O', `U' and `V' are shown in Figure~\ref{fig:box_plot_zeta_L_band}. 

It can be observed that the median value of $\zeta$ over `O' lies around \ang{29}. However, the standard deviation of $\zeta$ for the L band is much smaller than the C band. This phenomenon might be due to the longer wavelength of the L band, for which the ocean surface appears smoother concerning the C band. Also, over the urban area (`U'), we observe a decrease of $\approx$ \SI{15}{\percent} of the median value of $\zeta$. The decrease in $\zeta$ value might be due to the high penetration of the L-band for which it sensed more percentage of the ground component. A similar result is also evident over `V'. The median value of $\zeta$ over `V' is around \ang{37} while the standard deviation value is \SI{10}{\percent} lesser than C band data.

\section{Conclusions}
This study characterizes different canonical and natural targets on rotation in Euclidean space using full polarimetric Single Look Complex (SLC) data. In this regard, a new parameter, $\zeta$, is proposed, which can measure the oscillation of a target in the rotational domain. We observed that some targets which can be related through a unitary transformation with each other have the same $\zeta$ values. The variation of this parameter is then shown using the full polarimetric C band Radarsat-2 and L band ALOS PALSAR SLC data over the San Francisco Bay area and Cuba, respectively. Interesting differences are profound between C and L band wavelengths due to the different penetration capabilities.

\section{Acknowledgement}
Dr.~Dey would like to thank the Alexander von Humboldt Foundation for providing the ``Carl Friedrich von Siemens-Forschungsstipendium der Alexander von Humboldt-Stiftung'' fellowship to conduct the research at the German Aerospace Center (DLR), Germany. The authors thank MAXAR Technologies Ltd., formerly MacDonald, Dettwiler, and Associates (MDA), for providing RADARSAT-2 data and JAXA for the ALOS PALSAR data.

\small{
\bibliographystyle{IEEEtran}
\bibliography{bibliography}
}

\end{document}